\documentstyle[11pt,moriond,epsfig]{article}

\def\Journal#1#2#3#4{{#1} {\bf #2}, #3 (#4)}

\def\NIMA{{\em Nucl. Instrum. Methods} A}

\def\PLB{{\em Phys. Lett.}  B}

\def\be{\begin{equation}}
\def\ee{\end{equation}}
\def\bea{\begin{eqnarray}}
\def\eea{\end{eqnarray}}

\begin{document}
\vspace*{4cm}
\title{THE ONSET OF THE ANOMALOUS $J/\Psi$ SUPPRESSION IN Pb-Pb
COLLISIONS AT THE CERN SPS}

\author{ R. ARNALDI - NA50 collaboration  
\\
\vspace{0.3cm}
M.C.~Abreu$^{6,a}$,
B.~Alessandro$^{10}$,
C.~Alexa$^{3}$,
R.~Arnaldi$^{10}$,
M.~Atayan$^{12}$,
C.~Baglin$^{1}$,
A.~Baldit$^{2}$,
M.~Bedjidian$^{11}$,
S.~Beol\`e$^{10}$,
V.~Boldea$^{3}$,
P.~Bordalo$^{6,b}$,
S.R.~Borenstein$^{9,c}$,
G.~Borges$^{6}$
A.~Bussi\`ere$^{1}$,
L.~Capelli$^{11}$,
J.~Castor$^{2}$,
C.~Castanier$^{2}$,
B.~Chaurand$^{9}$,
B.~Cheynis$^{11}$,
E.~Chiavassa$^{10}$,
C.~Cical\`o$^{4}$,
T.~Claudino$^{6}$,
M.P.~Comets$^{8}$,
N.~Constans$^{9}$,
S.~Constantinescu$^{3}$,
P.~Cortese$^{10,d}$,
J.~Cruz$^{6}$,
N.~De Marco$^{10}$,
A.~De Falco$^{4}$,
G.~Dellacasa$^{10,d}$,
A.~Devaux$^{2}$,
S.~Dita$^{3}$,
O.~Drapier$^{11}$,
B.~Espagnon$^{2}$,
J.~Fargeix$^{2}$,
P.~Force$^{2}$,
M.~Gallio$^{10}$,
Y.K.~Gavrilov$^{7}$,
C.~Gerschel$^{8}$,
P.~Giubellino$^{10}$,
M.B.~Golubeva$^{7}$,
M.~Gonin$^{9}$,
A.A.~Grigorian$^{12}$,
S.~Grigorian$^{12}$,
J.Y.~Grossiord$^{11}$,
F.F.~Guber$^{7}$,
A.~Guichard$^{11}$,
H.~Gulkanyan$^{12}$,
R.~Hakobyan$^{12}$,
R.~Haroutunian$^{11}$,
M.~Idzik$^{10,e}$,
D.~Jouan$^{8}$,
T.L.~Karavitcheva$^{7}$,
L.~Kluberg$^{9}$,
A.B.~Kurepin$^{7}$,
Y.~Le Bornec$^{8}$,
C.~Louren\c co$^{5}$,
P.~Macciotta$^{4}$,
M.~Mac~Cormick$^{8}$,
A.~Marzari-Chiesa$^{10}$,
M.~Masera$^{10}$,
A.~Masoni$^{4}$,
M.~Monteno$^{10}$,
A.~Musso$^{10}$,
P.~Petiau$^{9}$,
A.~Piccotti$^{10}$,
J.R.~Pizzi$^{11}$,
W.~Prado da Silva$^{10,f}$,
F.~Prino$^{10}$,
G.~Puddu$^{4}$,
C.~Quintans$^{6}$,
S.~Ramos$^{6,b}$,
L.~Ramello$^{10,d}$,
P.~Rato Mendes$^{6}$,
L.~Riccati$^{10}$,
A.~Romana$^{9}$,
H.~Santos$^{6}$,
P.~Saturnini$^{2}$,
E.~Scalas$^{10,d}$,
E.~Scomparin$^{10}$
S.~Serci$^{4}$,
R.~Shahoyan$^{6,g}$,
F.~Sigaudo$^{10}$,
S.~Silva$^{6}$,
M.~Sitta$^{10,d}$,
P.~Sonderegger$^{5,b}$,
X.~Tarrago$^{8}$,
N.S.~Topilskaya$^{7}$,
G.L.~Usai$^{4}$,
E.~Vercellin$^{10}$,
L.~Villatte$^{8}$,
N.~Willis$^{8}$.
}
\address{\scriptsize{$^{~1}$ LAPP, CNRS-IN2P3, Annecy-le-Vieux,  France.
$^{~2}$ LPC, Univ. Blaise Pascal and CNRS-IN2P3, Aubi\`ere, France.
$^{~3}$ IFA, Bucharest, Romania.
$^{~4}$ Universit\`a di Cagliari/INFN, Cagliari, Italy.
$^{~5}$ CERN, Geneva, Switzerland.
$^{~6}$ LIP, Lisbon, Portugal.
$^{~7}$ INR, Moscow, Russia.
$^{~8}$ IPN, Univ. de Paris-Sud and CNRS-IN2P3, Orsay, France.
$^{~9}$ LPNHE, Ecole Polytechnique and CNRS-IN2P3, Palaiseau, France.
$^{10}$ Universit\`a di Torino/INFN, Torino, Italy.
$^{11}$ IPN, Univ. Claude Bernard Lyon-I and CNRS-IN2P3, Villeurbanne,
France.
$^{12}$ YerPhI, Yerevan, Armenia.\\
a) also at UCEH, Universidade de Algarve, Faro, Portugal
b) also at IST, Universidade T\'ecnica de Lisboa, Lisbon, Portugal
c) on leave of absence from York College CUNY
d) Universit\'a del Piemonte Orientale, Alessandria and INFN-Torino,
Italy.
e) also at Faculty of Physics and Nuclear Techniques, Academy of
   Mining and Metallurgy, Cracow, Poland
f) now at UERJ, Rio de Janeiro, Brazil
g) on leave of absence of YerPhI, Yerevan, Armenia
}}

\maketitle\abstracts{
The $J/\psi$ suppression observed by the NA50 experiment is one of the most 
striking signatures for quark gluon plasma formation in Pb-Pb collisions at 158 AGeV.
The $J/\psi$ production has been studied as a function of the centrality of the
collision estimated via the forward energy $E_{ZDC}$ released in a zero degree 
calorimeter (ZDC). The study of the correlation between the number of participant nucleons in the collisions, $N_{part}$, and 
$E_{ZDC}$ allows to check whether the $J/\psi$ suppression pattern vs. $E_{ZDC}$ is compatible with a sudden 
$J/\psi$ suppression mechanism expressed as a function of $N_{part}$. 
}

\section{Introduction}

If extreme conditions of high energy density and temperature are reached, a phase transition 
between hadronic matter and a state of deconfined quarks and gluons (QGP) should be achieved.

Experimentally, the best tool to study the formation of the QGP is provided by ultrarelativistic 
heavy ion collisions and several probes have been investigated as signatures of the phase 
transition. One of the most
striking signals is the $J/\psi$ suppression, as proposed by Matsui and Satz in 1986 \cite{Mat86}.
In fact, QGP formation induces a colour screening of the charmonium binding potential, 
preventing the c and $\bar{c}$ quarks to form charmonia bound states; hence a reduction of the $J/\psi$ 
production yield, the so-called $J/\psi$ suppression, is expected.

The NA38 experiment studied the $J/\psi$ production with p and S
induced reactions \cite{Abr98,Abr99,Abr992}. 

In such collisions, even if the required conditions 
for the phase transition were not achieved, a reduction of the $J/\psi$ production yield was 
already observed.
Such a ``suppression'' has been interpreted as due to nuclear absorption of the $c\bar{c}$
pair, before it forms a $J/\psi$, and it can be used as a baseline for the interpretation of the results
obtained by the NA50 experiment in Pb-Pb collisions.

NA50 has already published data on $J/\psi$ production as a function of the centrality estimator 
$E_T$ \cite{Abr993,Abr00}, i.e. the neutral transverse energy. 
The result indicates an additional $J/\psi$ suppression mechanism (the ``anomalous $J/\psi$ 
suppression''), not present in S--U interactions. 
The observed $J/\psi$ suppression pattern from peripheral to central collisions shows a two-step
behaviour, possibly linked with the successive melting in a deconfined state of the $\chi_c$  
which, through its radiative decay, is an important $J/\psi$ source, and of the strongly bound 
$J/\psi$ itself. A similar analysis performed as a function of another centrality estimator, $E_{ZDC}$, is presented
in this talk.

\section{The experimental apparatus and data taking conditions}

A detailed description of the NA50 apparatus can be found in \cite{Abr970,Arn98,Bel97}. 
The main component of the set-up is a dimuon spectrometer covering the pseudorapidity range 
$2.8<\eta<4.0$. 
Furthermore, NA50 is equipped with three 
centrality detectors: an electromagnetic calorimeter, measuring the neutral transverse energy 
$E_T$, a silicon microstrip detector, which allows to estimate the charged multiplicity and a 
zero degree calorimeter (ZDC). 
The ZDC, which covers the pseudorapidity region $\eta\geq 6.3$, is placed on the beam line, 
measuring the forward energy $E_{ZDC}$, mainly carried by projectile nucleons which have not 
taken part in the collision. 

The results discussed here refer to the 1996 and 1998 data taking periods. 
The differences between the two set-ups concern only the target region. 
In 1996, 7 sub-targets were used, with a total thickness corresponding to 30\% $\lambda_I$, 
while in 1998 only one single thin target (7 \% $\lambda_I$) was used. 
%For both years the average beam intensity was about 5$\cdot$10$^7$ Pb ions/burst, with a 5 s spill.

Data have been collected with two different triggers: the ``dimuon trigger'' and the ``minimum 
bias'' one.
The first one corresponds to events where the spectrometer detects a couple of muons produced in 
the target region, while the second one fires every time a small amount of energy is released 
in the ZDC. 
Information on data taking conditions and event selection can be found in \cite{Abr993,Abr00}.

\section{$E_{ZDC}$ and the centrality of the collision}

The geometry of the collision is usually characterized by the impact parameter {\em b}.
This variable is not directly measured, but it is accessible via measurements of other
quantities, like the zero degree energy $E_{ZDC}$. 
In fact, since the ZDC is placed on the beam line, it intercepts all the nucleons which
have not taken part in the collisions (i.e. spectator nucleons $N_{spect}$) and which fly towards the
detector, with the beam energy. Hence the measured zero degree energy is directly proportional 
to the number of spectator nucleons impinging on the detector and therefore to the impact
parameter {\em b}. 
%A small amount of $E_{ZDC}$ corresponds to a small number of spectator nucleons
%and hence to a small impact parameter. On the contrary a large amount of $E_{ZDC}$
%corresponds to a large number of spectator nucleons and therefore to a peripheral
%collisions.

%From the previous considerations we can express, for a generic Pb--Pb collision, the
%relation between the average $E_{ZDC}$ and {\em b} as:
%\begin{equation}  
%\label{eq:1}
%\centering  
%\langle E_{\rm ZDC}(b)\rangle =158 \times N_{\rm spec}(b) + \alpha\times 
%N_{\rm part}(b)=158 \times \left (208 - \frac{N_{\rm part}(b)}{2} \right) + 
%\alpha\times N_{\rm part}(b)
%\label{eq:ezdcvsb}
%\end{equation}      
For a generic Pb--Pb collision, the average $<E_{ZDC}>$ is given by the sum of two terms: a 
dominant one which refers to the contributions of the spectator nucleons, plus a smaller 
contribution proportional to the number of participant nucleons $N_{part}$ emitted in the 
calorimeter acceptance. The link between $N_{part}$ (or $N_{spec}$) and {\em b} has been obtained 
with a calculation based on a Glauber model of nucleus--nucleus collisions, using Woods-Saxon 
nuclear density profiles. 

For a generic {\em b} the $E_{ZDC}$ values are gaussian distributed around $<E_{ZDC}>$ with a 
width $\sigma_{E_{ZDC}}$  which takes into account both the finite resolution of the detector and 
the size of the physics fluctuations due to the width of the correlation between {\em b} and 
$N_{part}$.

\section{Study of the $J/\psi$ suppression}

\begin{figure}[ht]
\centering
\resizebox{0.4\textwidth}{!}{%
\includegraphics*[bb= 16 144 548 674]{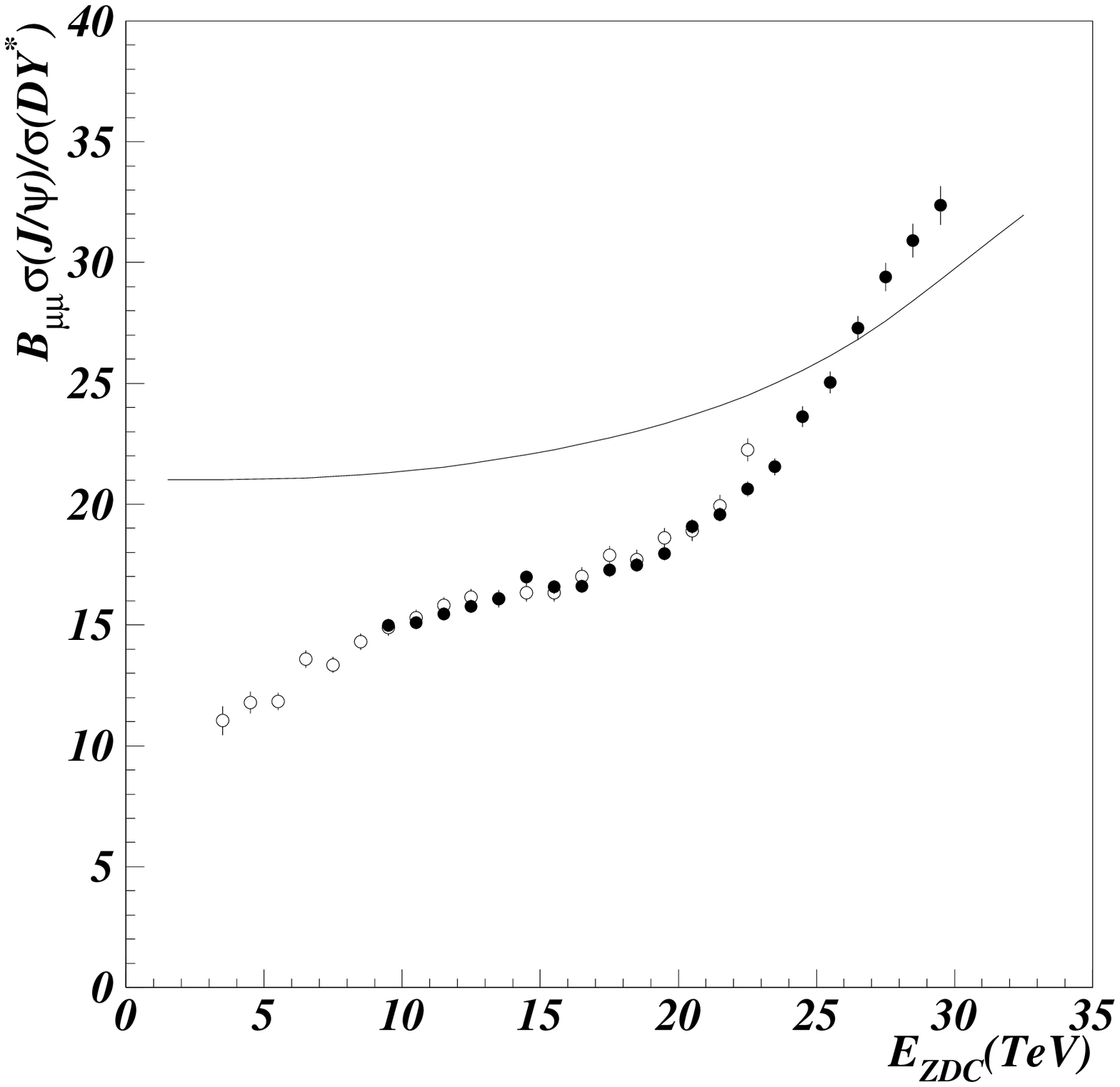}}
\resizebox{0.4\textwidth}{!}{%
\includegraphics*[bb= 20 148 535 659]{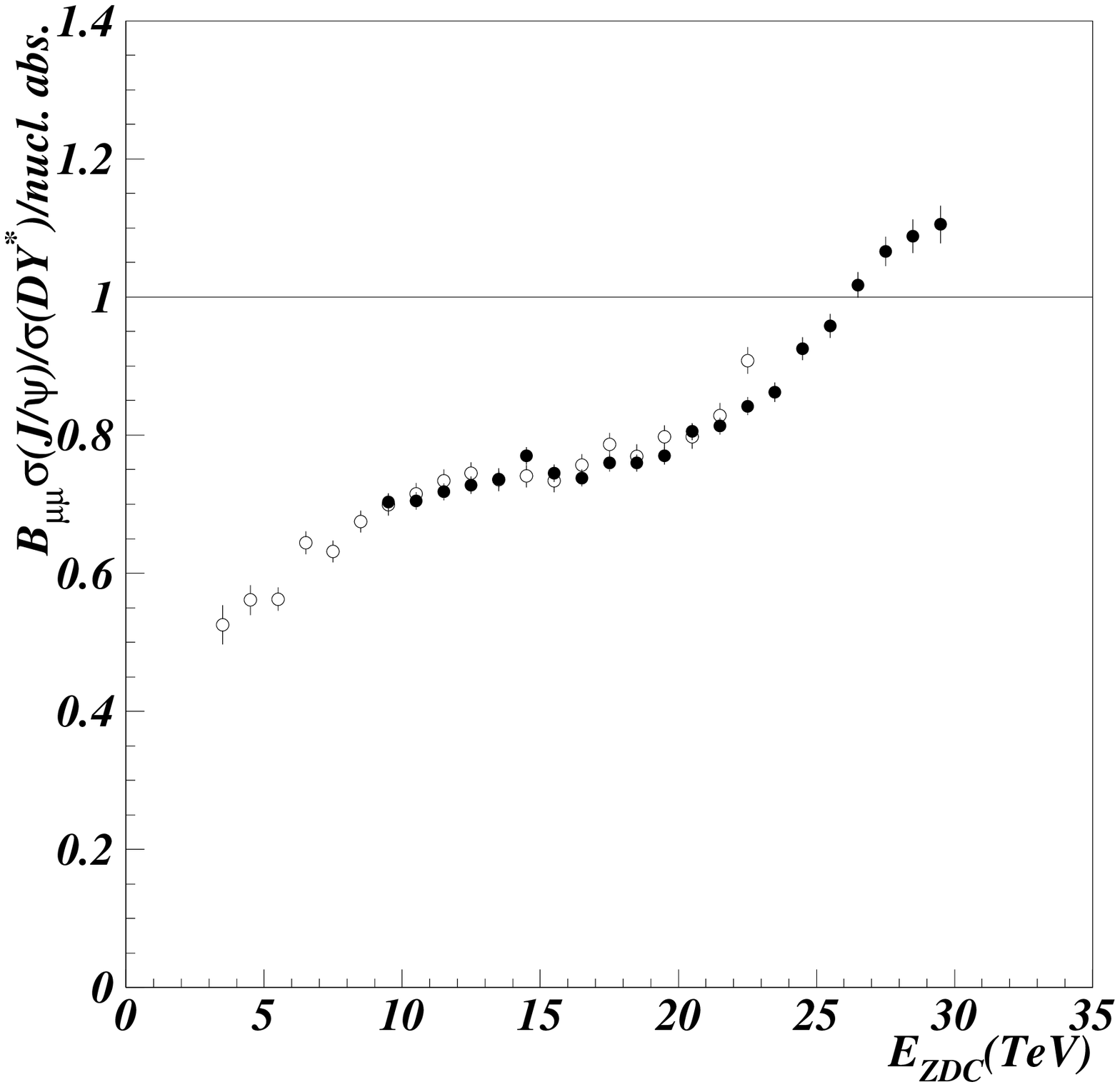}}            
\caption{(a) $J/\psi$ suppression as a function of $E_{ZDC}$. Full symbols refer to 1996
data and open symbols to 1998 data. The continuous line represents the $J/\psi$
suppression due to ordinary nuclear absorption; (b) the $J/\psi/DY^*$ ratio divided by the 
absorption curve, as a function of $E_{ZDC}$}
\label{fig:minbias}
\end{figure}

The $J/\psi$ suppression pattern as a function of $E_{ZDC}$ has been obtained with the
``minimum bias'' analysis. The explanation of this technique can be found in \cite{Abr993,Abr00}.
Basically the number of $J/\psi$ events, obtained from the $\mu^+\mu^-$ invariant mass spectrum, 
is divided by a ``calculated Drell--Yan ($DY^*$)'', estimated from the minimum bias sample of events.
The result is shown in Fig. \ref{fig:minbias}a, where the continuous line refers to the baseline
established by the data collected with lighter projectiles. 
Since 1996 and 1998 set--ups were respectively optimized for the study of peripheral and central collision, the results
are presented for the corresponding range of centralities.
In Fig. \ref{fig:minbias}b, the $J/\psi/DY^*$ result is divided by the nuclear absorption curve.

We observe a departure from the
expected standard behaviour around $E_{ZDC}\sim27$ TeV, while a second steep
decrease is visible in the region corresponding to the most central collisions.
A suppression pattern with the same characteristics has been obtained with the
analysis performed as a function of $E_T$ \cite{Abr993,Abr00}.

\section{$J/\psi$ suppression versus $N_{part}$}

%mettere nelle concl;usioni
%As discussed in \cite{Satz00}, one basic feature of the suppression pattern in a
%deconfinement scenario is a well defined onset. In fact all other approaches based on
%conventional suppression mechanisms predict a smoother trend as a function of
%centrality. 
The variable governing the onset of the anomalous suppression is a priori not
known. Hence, exploiting the link between $E_{ZDC}$ and $N_{part}$, we check if the  
suppression pattern of Fig. \ref{fig:minbias}b is compatible with two sharp drops in the 
$J/\psi$ yield occurring at definite values of $N_{part}$. In the interpretation of Ref.
\cite{Abr00} the two steps should correspond to the melting, in a deconfined state, of the 
$\chi_c$ with suppression of the $J/\psi$ from the decay $\chi_c\rightarrow {\rm J}/\psi\,\gamma$ 
at $N_{part}$=$N_1$, followed by the suppression of directly produced $J/\psi$ at $N_{part}$=$N_2$.
Taking into account the $N_{part}$ versus $E_{ZDC}$ 
correlation, and the finite resolution on $N_{part}$ due to the detector response, we calculate
the theoretical $(J/\psi/DY^*)/$Absorption ratio vs $E_{ZDC}$.

\begin{figure}[htbp]  
\centering  
\resizebox{0.4\textwidth}{!}            
{\includegraphics*[bb= 20 148 535 659]{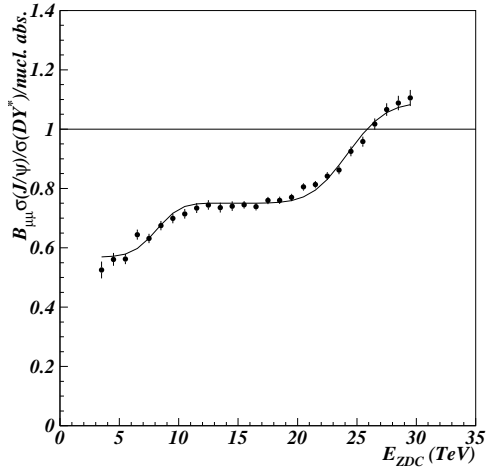}}            
\caption{The fit of $J/\psi/{DY^*}$ 
vs $E_{ZDC}$, assuming two sharp absorption mechanisms of free amplitude occurring at $N_1$=122 and $N_2$=334.}            
\label{fig:npart}            
\end{figure}

The agreement between the experimental data and this naive model is shown in Fig.
\ref{fig:npart}; the experimental points result to be well described with $N_1$=122, $N_2$=334. 
 
However, a more detailed study \cite{Arnaldi} shows that data could accomodate equally well an onset of the suppression smeared over a
certain limited $N_{part}$ range, not larger than 25 $N_{part}$.  

\section{Conclusions}
The $J/\psi$ suppression pattern as a function of $E_{ZDC}$ has been presented. 
The double step pattern, already observed in a similar analysis as 
a function of $E_T$ and interpreted as an evidence for deconfinement at SPS energy \cite{Abr00}, 
is found again in this analysis. 

Using the relation between $E_{ZDC}$ and $N_{part}$ the suppression pattern can be plotted
as a function of $N_{part}$. Taking into account the finite resolution on $N_{part}$ induced by
the experimental resolution on $E_{ZDC}$, it turns out that the onset of the anomalous $J/\psi$ suppression occurs
in a very limited $N_{part}$ range, smaller than 25 participant nucleons.

\end{document}